\documentstyle[twocolumn,aps]{revtex}
\begin{document}

\newcommand{\beq}{\begin{equation}}
\newcommand{\eeq}{\end{equation}}
\newcommand{\bea}{\begin{eqnarray}}
\newcommand{\eea}{\end{eqnarray}}
\newcommand{\ba}{\begin{array}}
\newcommand{\ea}{\end{array}}
\newcommand{\om}{(\omega )}
\newcommand{\bef}{\begin{figure}}
\newcommand{\eef}{\end{figure}}
\newcommand{\leg}[1]{\caption{\protect\rm{\protect\footnotesize{#1}}}}

\newcommand{\ew}[1]{\langle{#1}\rangle}
\newcommand{\be}[1]{\mid\!{#1}\!\mid}
\newcommand{\no}{\nonumber}
\newcommand{\etal}{{\em et~al }}
\newcommand{\geff}{g_{\mbox{\it{\scriptsize{eff}}}}}
\newcommand{\da}[1]{{#1}^\dagger}
\newcommand{\cf}{{\it cf.\/}\ }
\newcommand{\ie}{{\it i.e.\/}\ }
\newcommand{\eg}{{\it e.g.\/}\ }

\title{Reconstructing the formalism of quantum mechanics \\
in the  ``contextual objectivity"  point of view.}
\author{Philippe Grangier}
\address{Laboratoire Charles Fabry, Institut d'Optique Th\'eorique et
Appliqu\'ee, 
F-91403 Orsay, France}

\maketitle

\begin{abstract}

In a previous preprint \cite{ph1} we introduced a ``contextual objectivity" 
formulation of quantum mechanics (QM). A central feature of this approach is
to define the quantum state in physical rather than in mathematical terms,
in such a way that it may be given an ``objective reality". 
Here we use some ideas about the system dimensionality, taken from \cite{lh}, 
to propose a possible axiomatic approach to QM.
In this approach the structure of QM appears as a direct consequence
of the non-commutative character of the (classical geometrical) 
group of ``knobs transformations", that relate between themselves
the different positions of the measurement apparatus.



\end{abstract}

\section{Introduction}

In a previous preprint \cite{ph1}, we introduced and discussed a ``physical"
(as opposed to mathematical)  definition of  
a quantum state\footnote{Throughout this paper ``state" means ``pure state". 
Mixed states, when needed, will be called ``statistical mixtures" \cite{cct}.}, 
that reads in the following way : 

{\bf The quantum state of a physical system is defined by the values of a complete
set of physical quantities, which can be predicted with certainty and measured
repeatedly without perturbing in any way the system} (the set of quantities is
complete in the sense that the value of any other quantity which satisfies
the same criteria is a function of the set values).

As discussed in detail in  ref. \cite{ph1}, and briefly recalled below, 
this definition is in full agreement with the usual formalism of QM
(it was actually deduced from it). But here 
we would like to address another question : is possible to {\bf deduce} the usual
formalism of QM from this definition ? In order to address this question, 
we will discuss a procedure that was introduced by Lucien
Hardy in \cite{lh}. Then we will introduce an alternative approach, that 
differs significantly from \cite{lh}, since it is based upon geometry rather
than upon probabilities. 

\section{A few words about contextual objectivity}

The definition of the quantum state that is given 
above is clearly in agreement with the usual formalism of
QM, as it can be seen when using the notion of ``complete set of
commuting observables" (CSCO) \cite{cct}. A quantum state is specified by the
ensemble of eigenvalues corresponding to a CSCO, that can obviously be
measured repeatedly without perturbing in any way the system.
Actually, as said in the definition, 
any physical quantity that satisfies the definition can be
expressed as a function of the CSCO ones \cite{ph1}.
This also appears in the sentence ``measured
repeatedly without perturbing in any way the system": if the subset is not a
CSCO, a measurement will generally change the state. 
It should be clear also that unitary 
evolution from Schroedinger's equation transforms a state which satisfies our
definition into a similar state, associated with a different set of physical
quantities, corresponding to new well-defined measurements 
(that may however be not easy to perform).

An obvious but fundamental point is that given a quantum
state, not all possible physical quantities can be predicted with certainty, but
only a subset of them. That the subset is the largest 
possible set of independant quantities is just the
definition of a CSCO. How to deal with
physical observables which do not commute with those of the CSCO is a
crucial point, that is discussed in detail in \cite{ph1}.
It is  also argued in \cite{ph1} that within this framework, there is no need
to add a ``measurement postulate": this postulate is already contained in the
definition of the quantum state,
\ie, the very possibility to define a quantum state and the measurement postulate
are essentially identical. For more details about that, including how it relates
to the EPR ``paradox" \cite{epr} or to the decoherence approach, we refer the reader to \cite{ph1}.

It is worth pointing out that our definition
implies that some ``objectivity" can be attached to the quantum state. 
This is because the quantum state is associated
with a fully predictable course of events, that is independent of the observer. 
Actually, our definition assumes that
the sentence ``predicted with certainty and measured repeatedly" itself has a
meaning, \ie, that we are able to make experiments that correspond to 
well-defined measurements, and that give predictable outcomes. 
This is why we call our
point of view ``contextual objectivity": the quantum state does have an
objective existence, but its definition is inferred from observations that are
made {\bf at the macroscopic level}. We point out that there is no need to refer to
``observer's consciousness" or anything like that: what we need is
simply the usual classical world, as the place where the measurements are made, and
where results can be recorded by any (conscious or unconscious) observer.
How exactly the ``quantum reality" is related with the ``macroscopic reality"
will be spelled out in the conclusion section of this paper.

\section{Exclusive and non-exclusive modalities}
\label{meas}

We shall now forget about the QM formalism, and look how far we can go by using
only our definition of a quantum state. What we have at  hand is only
pure states, that yield predictions with unit probability for a given
set of measurements. This can be detailed further :

(i) for a given set of (orthogonal) measurements, the 
different pure states corresponding to different results
are exclusive, and we will call the associated pure states
``exclusive modalities". We may assume that the number of exclusive
modalities is a property of the system, that will be called the dimension, $N$.

(ii) from our definition of a quantum state, there are other sets of
measurements that are not predicted with certainty if the state of the
system is one of the $N$ exclusive modalities. It is thus natural to
consider all possible pure states, that constitute 
``non-exclusive modalities" (using the usual terminology for clarity, 
they are eigenvectors  of other observables that do not commute with the ones in the initial
set, but the QM formalism is not required at that stage).

The existence of non-exclusive modalities is a specific
quantum feature : in classical physics, it should be possible to 
``give more details" about the state (\eg by increasing the number of measurements),
so that the ``fully defined" states are exclusive,
but this is not possible in QM. This is contained in our definition 
of the quantum state, and corresponds physically both to the existence of
Heisenberg inequalities, and to the fact that a pure quantum state
cannot be ``completed" \cite{epr}. This specifically quantum difference between 
exclusive and non-exclusive modalities will be very important in the following.

One may thus ask the question : is it possible to reconstruct the QM formalism 
by appropriately combining exclusive and non-exclusive modalities ? 
It should be clear that if one restricts to
exclusive modalities, one will obtain something very close to classical
probability theory over a discrete set of $N$ exclusive events, 
but there will be no room for interference effects, and thus we will miss QM.
Lucien Hardy proposed in \cite{lh} an approach based on 
using $K$ probabilities defined on non-exclusive modalities, in such a way that
they define an arbitrary (pure or mixed) quantum state. This approach
is discussed in some details in the Appendix, and as a conclusion of this
discussion we argue that the required axioms on $K$ are difficult to justify.
In the following we will thus present another approach, that does not
require the axioms on $K$ (H2 and H4b, see Appendix).
On the other hand, we essentially keep H1 (statistical interpretation
of probabilities), H3 and H4a (definition and properties of $N$),
and H5 (continuity). As a central difference with Hardy's approach,
we will consider this continuity axiom from a geometrical point of view.
This will lead us to the conclusion that QM is not only a probability theory, 
but has also to deal with geometry as an essential ingredient.

\section{Reconstructing quantum mechanics}

\subsection{Restating the question}

The question that we want to address is : since we know that there are $N$ exclusive modalities 
associated to each given CSCO, how to connect between themselves all the non-exclusive modalities
corresponding to all possible CSCO ?  We should emphasize that the notion of Hilbert space
is not yet there : beyond the axioms themselves, what we have is simply
our definition of a quantum state, as the values of a
set of physical quantities, that can be predicted with certainty and measured repeatedly.

By definition, changing the CSCO results from changing the measurement apparatus 
at the macroscopic level, that is, ``turning the knobs". A typical example is changing
the orientations of a Stern-Gerlach magnet. These transformations have the
mathematical structure of a continuous
group: the combination of several transformation is associative 
and gives a new transformation,
there is a neutral element (doing nothing), and each transformation has an inverse. 
Generally this group is not commutative : for instance, the three-dimentional
rotations associated with the orientations of a Stern-Gerlach magnet do not commute.
For a given position of the knob settings (given CSCO), there is a given set
of $N$ exclusive modalities, that we can denote $\{b_i\}$. By turning the knobs,
one obtains $N$ other exclusive modalities, that we can denote $\{b_j'\}$.
The question is then how to relate the  $\{b_i\}$ and the $\{b_j'\}$, that is, 
to define the transformation that allows one to go from a set of states of the other one.
The reasoning presented below is not a demonstration, but a justification 
of a ``reasonable" way to establish this relation.

\subsection{The probability formula}

Since the $\{b_i\}$ and $\{b_j'\}$ are by definition non-exclusive modalities,
one has first to introduce the probabilities of finding the particular state
$b_j'$ (after ``turning the knobs"), when one starts in state $b_i$ (before ``turning the knobs").
Since there are $N^2$ such probabilities,
one can arrange them in a matrix $\Pi = \left( p_{i,j} \right)$, corresponding
to all the probabilities
connecting the two sets  $\{b_i\}$ and $\{b_j'\}$. Due to normalization conditions, it is
easy to check that the number of independant numbers in $\Pi$ is equal
to $(N-1)^2$ (these numbers have to fulfill appropriate inequalities for consistency). 
In order to manipulate the $\Pi$ matrix, it is convenient to introduce 
the orthogonal ($N \times N$) projectors $P_i$, with $P_i P_j = P_i \delta_{ij}$.
A useful operation is then to extract the particular probability $p_{i,j} $ from the matrix $\Pi$. 
It is easy to check that one may write:
\beq
p_{i,j} = Trace(P_i \;\;  \Sigma \;  \;  P_j \; \;  ^t\Sigma)
\eeq
where $\Sigma = \left( \sqrt{p_{i,j}} \right)$
is the matrix formed by the square roots of the probabilities, and $^t$ denotes the 
transpose operation. 
Since we are looking for a state transformation, 
and since the $\{P_i\}$ define a set of orthogonal projectors 
corresponding to the initial set of states, 
it is very tempting to consider that 
the relations $P_j' = \Sigma \; \; P_j  \; \; ^t\Sigma$ define the set of projectors corresponding
to the transformed set of states $\{b_j'\}$ . However, though
the $P_j'$ are indeed projectors (because the diagonal terms of $^t\Sigma \; \Sigma$
are all equal to 1), they are not {\it orthogonal} projectors
(because the off-diagonal terms of $^t\Sigma \; \Sigma$ are not zero). 
The solution to that problem is obvious :
the real-numbers matrix $\Sigma$ has to be replaced by the matrix 
$\tilde\Sigma = \left( e^{i \phi_{i,j}} \sqrt{p_{i,j}} \right)$,
where the phase factors $ e^{i \phi_{i,j}} $ are chosen so that $\tilde\Sigma$ is unitary
(the normalization conditions on the $p_{i,j}$ warrants that this is possible; note that 
complex numbers are required : $\tilde\Sigma$ cannot be simply an orthogonal matrix).
The equation for picking up a particular probability becomes:
\beq
p_{i,j} = Trace(P_i \;\;  \tilde\Sigma \; \; P_j  \; \; \tilde\Sigma^\dagger)
\label{equ}
\eeq
that is the expected formula. 
As a consequence, what we actually need is a set of $N \times N$ unitary matrices,
each one being associated to an element of the group of ``knobs transformation",
that will be denoted as ${\cal G_{K}}$.
For general consistency of the approach, we may conclude that
this set of matrices gives a representation of the group of knobs transformations;
this is fully consistent with the well known Wigner theorem \cite{fl}.
We note that we never used the definition of a state as a ``ray in an Hilbert space".
The state keeps its initial definition as a set of measurement results, but what we get
is a way to connect between themselves the probabilities 
of the ``non-exclusive modalities" associated to all possible pure states.
This connection is simply made by eq. (\ref{equ}), where the 
$\tilde \Sigma$ are unitary matrices that form a representation of the group
of ``knob transformations" ${\cal G_{K}}$.

It is noticeable that if ${\cal G_{K}}$ is commutative, then its $N \times N$ representations
are diagonal matrices with complex numbers of modulus unity of the diagonal, and zeros
everywhere else. These matrices commute with $P_j$, and thus $p_{i,j} = \delta_{i,j}$.
Therefore, there is actually no need for non-exclusive 
modalities, the initial $N$ exclusive ones are enough. This provides a straightforward
way to recover classical probability theory, and the mathematical structure of QM
thus appears as a direct consequence of the non-commutative character of ${\cal G_{K}}$ \cite{fl}.

\section{Conclusion}

As a summary, we sketched a possible way to reconstruct quantum mechanics
in the framework of the ``contextual objectivity" point of view. 
Like in \cite{lh}, we use an axiom on the system's dimension, that is
related to a well-defined number of ``exclusive modalities" for a 
given quantum system. Then we use geometry to connect
the different ``exclusive modalities" corresponding to different
settings of the measurement apparatus.
Our reconstruction axioms can thus be written as : 

\begin{itemize}

\item {\bf Axiom 1 }: The quantum state of a physical system is defined as the values of a
complete set of physical quantities, that can be predicted with certainty and measured
repeatedly without perturbing in any way the system. In the axioms below a quantum state 
will be called a ``modality".

\item {\bf Axiom 2 }:  For a given ``knob settings" of the measurement apparatus,
there exist $N$ distinguishable states $\{b_i\}$, that are called ``exclusive modalities". 
The value of $N$, called the dimension, is a characteristic 
property of a given quantum system \cite{note2}.

\item {\bf Axiom 3 }:  Various knob settings are related between themselves 
by (classical geometrical) transformations $g$ that
have the structure of a continuous group ${\cal G_{K}}$. 

\item {\bf Axiom 4 (theorem ?) }: If the system is known to be in the state $b_i$ from the set $\{b_i\}$,
the probability that it is found in state $b_j'$ from the set $\{b_j'\}$ 
corresponding to another knob settings obtained by 
the knob transformation $g$ is : 
\beq
p_{i,j} = Trace(P_i \;\;  \tilde\Sigma \; \; P_j  \; \; \tilde\Sigma^\dagger)
\eeq
where the $P_i$ are orthogonal projectors, and $\tilde\Sigma$ is a unitary matrix
corresponding to $g$, 
within the $N \times N$ matrix representation of the group ${\cal G_{K}}$. 

\end{itemize}

Whether or not Axiom 4 can be seen as a theorem (similar to Gleason's theorem \cite{caf},
but with slightly different hypothesis) remains an open question.
It should be clear that 
the standard structure of QM can be obtained from the above axioms
(in particular rewriting physical quantities  as operators and states
as rays is straightforward). We note again that in our approach
there is no ``measurement postulate", since it is already included
in Axiom 1 (see detailed discussion in \cite{ph1}). 

An important feature of our approach is that the mathematical 
structure of QM is a direct consequence of the non-commutative
character of the group of knobs transformations ${\cal G_{K}}$. 
{\it In some sense, QM appears as the result of
accomodating the ``contradictory" requirements that 
the exclusive modalities have a discrete structure, 
and that the knobs transformations
have a continuous, but non-commutative group structure.}

The view about the ``classical vs quantum" dilemma 
that emerges from our approach is thus the following.
A physical quantity is defined as an ensemble of possible measurements,
that are connected between themselves by ``geometrical" transformations
that are in the  ``knob transformations" group ${\cal G_{K}}$. 
The ``classical illusion" is to identify this physical quantity with
the numbers given by the measurement, and to attribute ``reality" to these numbers .
EPR themselves realized that this definition of ``reality" was too
restrictive, and proposed instead their definition based upon
predictability and reproducibility; 
this is just the idea that we use as our definition
of a quantum state. But as soon as this is done, 
it appears that this ``reality" cannot be attributed simultaneously
to all physical quantities : this is simply incompatible with the structure of
${\cal G_{K}}$. Thus what is ``real" at the macroscopic level is 
the definition of the physical quantities (\ie of the possible measurements
related by the group ${\cal G_{K}}$), and 
what is ``real" at quantum level (\ie at the level of the measured system)
is the quantum state. These two ``realities"
are fully compatible - they are actually the only ones that can
connect the experimental definition of a physical quantity and the measurement results 
in a consistent way \cite{note3}.

\section*{Acknowledgements}

Stimulating discussions with Jean-Louis Basdevant, 
Franck Laloe and Anton Zeilinger are acknowledged, as well as
interesting email exchanges with David Deutsch,
Lucien Hardy and Chris Fuchs.

\section*{Appendix : Hardy's approach}

\subsection{The problem}

The main idea of this approach is to define an arbitrary state of the system (that may be
pure or not pure) by using a set of real numbers, each representing
the probability that the system is in a  given {\it pure} state, chosen among a fixed set of
non-exclusive modalities.  
We note that it is not warranted that such a procedure should be successful :
there is an infinite number of non-exclusive modalities, and no obvious reason
that all of them can be represented by using probabilities of being in 
a finite number of modalities taken from all possible ones.
Nevertheless, following Hardy \cite{lh}, we will assume that this can be done, 
and we will denote as K the number of probabilities needed to determine 
an arbitrary state, defined over  a set of K non-exclusive modalities.
The number K should be related to N, that is the number of exclusive modalities
(dimension of the system). 
Therefore the questions can be rewritten as :

- it is possible to find out the number K of non-exclusive modalities that are
required to define an arbitrary (pure on non pure) state of the system, by using
the probabilities of being in each of the modalities ?

- is it possible to recover the usual structure of QM from these hypothesis ?

The answers to these questions are yes, as it was shown by Hardy in \cite{lh},
provided that some simple axioms are introduced about the
structure of a probability theory. In particular, Hardy shows
that for classical probability theory one has $K = N$,
and for quantum probability theory one has $K
= N^2$ [this is how many real numbers are required to
define a non-normalized density matrix].

\subsection{``Five reasonable axioms"}

The probability axioms introduced by Hardy are the following (here they are slightly reworded
and  reordered, keeping the initial numbering) :

\begin{itemize}

\item Axiom on probabilities (H1) : Relative frequencies (measured by
taking the proportion of times a particular outcome is observed)
converge to the same value, which we call the probability, for any case
where a given measurement is performed on a ensemble of n systems
prepared by some given preparation in the limit as n becomes large. 

\item Axioms on N (H3 and H4a) : 
The dimension N is a property of a given system, related to the number
of exclusive modalities that can be defined for that system.
A system whose state is constrained to have support
on only M of a set of N possible exclusive modalities behaves like a
system of dimension M (H3). 
A composite system consisting of subsystems A and B satisfies $N = N_A N_B$ (H4a).

\item Axioms on K (H2 and H4b) :
An arbitrary state of the system can be defined by specifying probabilities 
over a set of K non-exclusive modalities. 
K is determined by a function of N (i.e. $K = K(N)$),
and for each given N, K takes the minimum 
value consistent with the other axioms (H2). 
A composite system consisting of subsystems
A and B satisfies $K = K_A K_B$ (H4b).

\item Axiom on pure states (modalities) :
The above axioms are consistent with both classical and quantum
probability theories. To get quantum theory one must add another
axiom about ``continuity" : any pure state can be transformed continuously
and reversibly along a path through the pure states to any other pure state (H5).

\end{itemize}

\subsection{Discussion}

The axioms H1, H3 and H4a about probabilities and about N are easy to accept,
since they deal with the definition of a probability theory over
an ensemble of N exclusive modalities. In particular, the arguments 
about subspaces and composite systems are natural once it is admitted that
the number of exclusive modalities is a characteristic property of the system.
Similarly, the ``quantum" axiom H5 is acceptable given the initial definition of
a pure state as a modality, since an ``infinesimal" change to the measurement set-up is
expected to change continously the resulting pure states. 

On the other hand, the axioms on K (H2 and H4b) are not obvious :  
they are trivial consequences of the previous ones 
if $K=N$ (that is the classical probability situation), but far
from trivial if $K \neq N$. 
Can our approach be helpful ?
Accepting the (non-obvious) step that K is a function of N only, we interpret
the equality $K = K_A K_B$  as meaning that the set of non-exclusive modalities
(\ie the K's) is treated just like the set of exclusive ones (\ie the N's) 
by the probability theory. This is consistent with considering the 
K states as ``real", though they are not exclusive of each other, but
still does not provide a straightforward justification for these axioms. 

The content of the axioms may be better understood by looking at their
implications. An immediate consequence
of H4a and H4b is that $K(N^2) = K(N)^2$. By adding some arguments to show
that K must be a polynomial in N \cite{lh}, this implies that  $K(N) = N^r$,
where $r$ is an integer. In a previous version (v1) of \cite{lh}
the value of $r$ was claimed to be calculated, but in the present version (v3) 
it is simply shown that $r$=2, \ie $K(N) = N^2$, is the smallest value
compatible with H5, and corresponds thus to quantum theory. 
If H5 is dropped, then $K = N$ is acceptable and classical probability theory is obtained.
After prooving that $K = N^2$, Ref. \cite{lh} shows how to reconstruct a ``Liouville space" 
(rather than Hilbert space) formulation of QM, 
including  the trace formula and the general structure of the time evolution. 

In our opinion, an important problem in the approach of \cite{lh} is that 
the axioms on K (H2 and H4b)  are difficult to justify. 
In particular, it seems unlikely
that (quoting \cite{lh}) ``a 19th century theorist may have
developed quantum theory without access to the empirical data that later
became available to his 20th century descendants". Without a
good knowledge about QM, there is little chance to reach the crucial idea of
``non-exclusive modalities" (non-orthogonal pure states as irreducible
features of the reality), that looks crucial for the whole reasoning.
Though our contextual objectivity approach may
help to clarify physically the ``K vs N" issue, it does
not provide a firm basis to justify all the axioms of Ref. \cite{lh}.
As we have shown above, the ``missing axioms" can advantageously be replaced
by geometrical considerations.


\begin{references}

\bibitem{ph1} Philippe Grangier, {\it ``Contextual objectivity : 
a realistic interpretation of quantum mechanics"}, arXiv: quant-ph/0012122

\bibitem{lh} Lucien Hardy, {\it ``Quantum Theory From Five Reasonable Axioms"}, 
arXiv: quant-ph/0101012

\bibitem{epr} A. Einstein, B. Podolsky and N. Rosen, {\it ``Can quantum mechanical
description of reality be considered complete"}, Phys. Rev. {\bf 47}, 777
(1935)

\bibitem{cct} C. Cohen-Tannoudji, B. Diu and F. Laloe, {\it ``M\'ecanique
Quantique"}, Hermann, 1977

\bibitem{fl} The group of geometrical displacements is a very important example, 
where $N$ is actually infinite. But the same considerations can be translated
to that case, see \eg, Franck Laloe, {\it ``Les sym\'etries en m\'ecanique quantique"},
Cours de DEA, Ecole Normale Sup\'erieure (1980). 

\bibitem{note2} The value of $N$ may be finite or infinite,
the important point is that there exist also  a (much larger) number of ``non-exclusive modalities"
(associated with all possible knob settings), that {\it cannot} be reduced to exclusive modalities.
For other discussions of extensions to infinite N, see also \eg \cite{lh} and \cite{fl}.

\bibitem{caf} C. A. Fuchs, 
{\it ``Quantum Foundations in the Light of Quantum Information"}, arXiv: quant-ph/0106166. 
It should be clear that we do  not share all positions expressed
in this reference, since our  goal is rather to deal with
``Quantum Information in the Light of Quantum Foundations".


\bibitem{note3} Among open questions, we did not consider time evolution. One should also
spell out in more detail the known connection between the physical
quantities and the infinitesimal generators
of ${\cal G_{K}}$. Another interesting
question is the role of ``projective" representations \cite{fl},
that are connected to gauge theories for systems that involve
charged particles and electromagnetic fields.

\end{references}
\end{document}